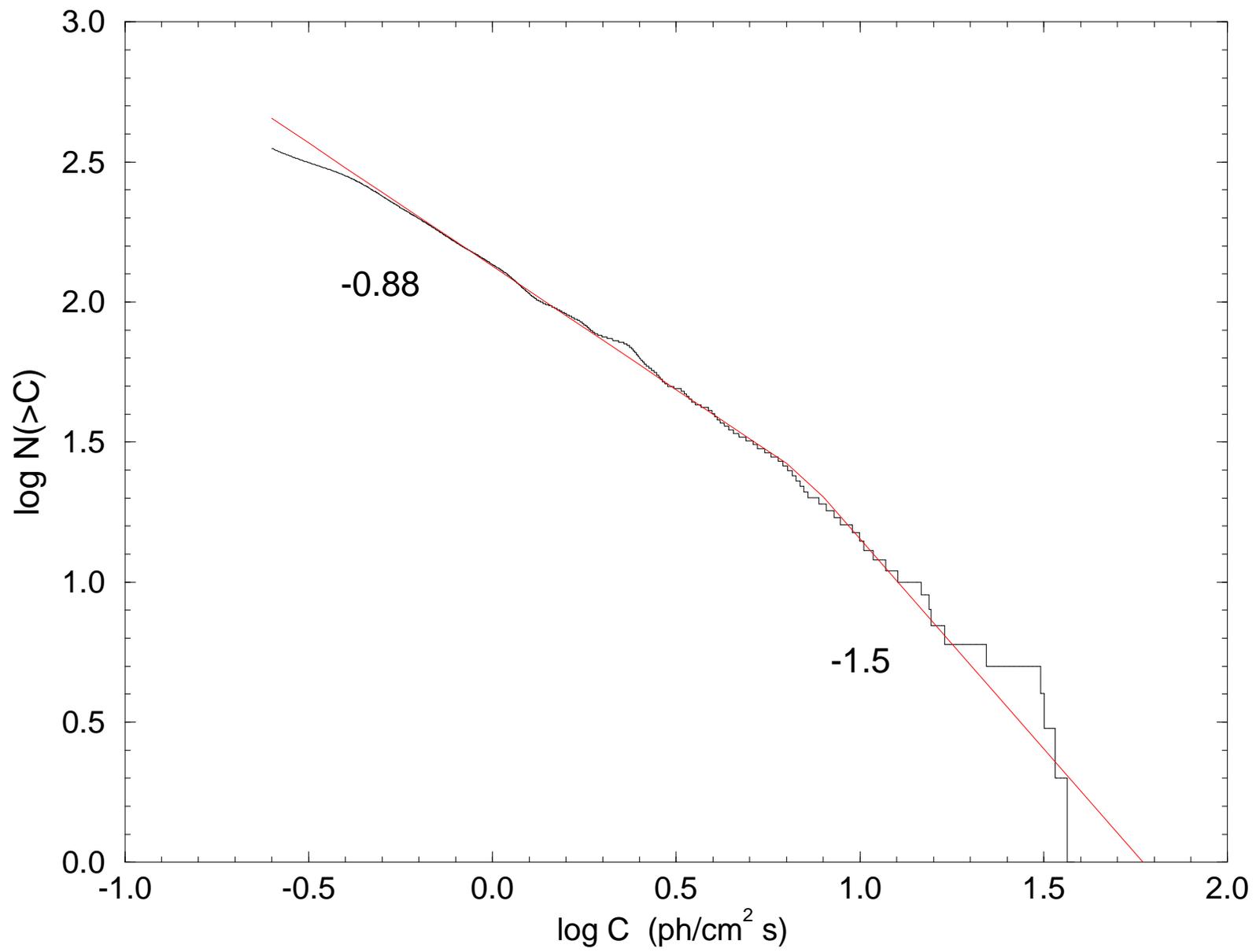

# The Brightness Distribution of Bursting Sources in Relativistic Cosmologies

P. Mészáros[1] and A. Mészáros[1,2]

[1]Pennsylvania State University, 525 Davey Laboratory, University Park, PA 16803, USA
[2]Department of Astronomy, Charles University, 150 00 Prague 5, Švédská 8, Czech Republic

## ABSTRACT

We present analytical solutions for the integral distribution of arbitrary bursting or steady source counts as a function of peak photon count rate within Friedmann cosmological models. We discuss both the standard candle and truncated power-law luminosity function cases with a power-law density evolution. While the analysis is quite general, the specific example discussed here is that of a cosmological gamma-ray burst distribution. These solutions show quantitatively the degree of dependence of the counts on the density and luminosity function parameters, as well as the the weak dependence on the closure parameter and the maximum redshift. An approximate comparison with the publicly available Compton Gamma Ray Observatory data gives an estimate of the maximum source luminosity and an upper limit to the minimum luminosity. We discuss possible ways of further constraining the various parameters.

## 1. INTRODUCTION

Gamma-ray burst (GRB) sources appear to be distributed isotropically down to the current limit of spatial resolution (Meegan et al. 1992, Fishman et al. 1994, Meegan et al. 1994), suggesting a cosmological origin. The brightness distribution, or the integral number counts of GRB as a function of peak photon count rates have been discussed in this context by, among others, Mao & Paczyński (1992), Dermer (1992), Piran (1992) and Wasserman (1992). Statistical fits to this "$\log N - \log C$" distribution have been carried out recently by Wickramasinghe et al. (1993), Horack, Emslie & Meegan (1994), Cohen & Piran (1995), and Emslie & Horack (1994). Some of the questions addressed in these references include the degree of sensitivity of the GRB count rates to the details of the cosmological model, the density evolution and the luminosity function of the bursts.

Except for the simpler standard candle case, these calculations have required extensive numerical integrations in order to explore the parameter space fully. These numerical calculations have shown that the fits are not very sensitive to the exact cosmological model, and that bounds on the density evolution and luminosity function are fairly general in nature. The role of each individual parameter in such numerical fits can only be inferred by varying them one at a time and recalculating the fit in each case. The analysis relies therefore on extensive sets of graphs and tables. A physical interpretation of the reason why the fits vary as they do as function of the different parameters would be greatly aided if one had analytic expressions for the quantities being fitted. These could, furthermore, greatly reduce the amount of numerical work involved, while making it easier to explore the effect of alternative distribution forms or evolution laws.

In the present paper, we derive closed form analytical solutions for the integral burst number count rates for different cosmological models, including a source density evolution proportional to a power of the expansion scale and a power-law luminosity distribution with upper and lower limits. These analytic expressions provide direct quantitative measures of the relative contributions of these three different physical factors, and show explicitly the reasons for which the comparisons of the models to the data are largely insensitive to the specific values of the closure parameter $\Omega_o$ and the density evolution power law index. The analytic solutions show explicitly the functional similarities and differences between a standard candle interpretation and a power-law luminosity distribution case. For a luminosity distribution, the analysis provides a direct estimate not only of the luminosity function slope, but also of the maximum burst luminosity, as well as lower limits for both the ratio of maximum to minimum burst luminosities and the total burst rate per galaxy. In this paper we discuss approximate fits-by-eye of the analytical solutions to the observed $\log N - \log C$ curves data, and the general behavior of the analytical solutions. These solutions are expected to be very useful in checking numerical integrations, and also in performing detailed statistical fits, which we intend to discuss elsewhere.

The analysis considered here is fairly general and may apply to other phenomena as well, e.g. supernova outburst rates in external galaxies, extragalactic gravitational wave or neutrino burst events, radio galaxy flare-ups, etc. These are also phenomena where luminosity function and evolutionary effects can play a large role, and are not treated here. It can also apply to steady (non-bursting) sources, if one replaces the parameter $D$ by $D_{ss} = D + 1$, where $D$ has the same meaning as in eq.(2). Our discussion here is developed for the specific case of a cosmological gamma-ray burst distribution.

## 2. BURST RATE AND PHOTON COUNT MEASUREMENTS

The number of generic transient events (henceforth bursts, or GRBs) observed per year out to some redshift $z$ can be calculated if one knows the local burst rate density as a function of redshift. Since one is interested in observable bursts, their intrinsic luminosity must be taken into account, as well as the fraction of it which is actually observable by a particular instrument. Consider a GRB with a peak bolometric photon emissivity $\mathcal{L}$ (photons per second) at a redshift $z$. A K-correction factor will take into account the spectrum of the source and the fact that the detector measures only within a limited band of frequencies $\nu_1$ and $\nu_2$ (Mao & Paczyński 1992). If the photon spectrum is characterized by a power law $\mathcal{L}_\nu \sim \nu^{\alpha-2}$, where $\alpha$ is the index for the power-per-decade spectrum below (generally different below and above several hundred keV, c.f. Higdon & Lingenfelter 1990), then the effective measured "bolometric" peak emission rate will be

$$\mathcal{L} = \int_{\nu_1(1+z)}^{\nu_2(1+z)} \mathcal{L}_\nu d\nu = (1+z)^{(\alpha-1)} \mathcal{L}_K \quad \mathcal{L}_K = \int_{\nu_1}^{\nu_2} \mathcal{L}_\nu d\nu. \quad (1)$$

For $\alpha \simeq 1$ (typical of GRB spectra below about 0.5 MeV, where most of the measured photons are (e.g., Band et al. (1993)), the K-correction is not present. In what follows, except for the short note in §2 below eq.(9) we will assume $\alpha = 1$ and $\mathcal{L}_K = \mathcal{L}$.

The typical photon emission rate $\mathcal{L}$ and the number density of bursts may depend on the redshift $z$. We denote by $\tilde{\Phi}(\mathcal{L}, z)d\mathcal{L}$ (in Mpc$^{-3}$ yr$^{-1}$) the number of GRB per year per unit proper volume with emissivity between $\mathcal{L}$ and $(\mathcal{L} + d\mathcal{L})$. This may be parametrized as

$$\tilde{\Phi}(\mathcal{L}, z) = \Phi(\mathcal{L})(1+z)^D, \quad 0 \leq D \leq 4, \quad (2)$$

where $\Phi(\mathcal{L}) = \tilde{\Phi}(\mathcal{L}, 0)$, and $(1+z)^D$ is used to define the density dependence. $D = 3$ would correspond to a constant comoving density, while for $D > 3$ ($D < 3$) the comoving density decreases (increases) with time. (The restriction on the values of $D$ is a matter of convenience only, other values being in principle allowable too. In this paper, for simplicity, we restrict ourselves to integer values $0 \leq D \leq 4$). The total number of GRB per Mpc$^3$ per year at redshift $z$ is then

$$n(z) = \int_{\mathcal{L}_{min}}^{\mathcal{L}_{max}} \tilde{\Phi}(\mathcal{L}, z) d\mathcal{L} = (1+z)^D \int_{\mathcal{L}_{min}}^{\mathcal{L}_{max}} \Phi(\mathcal{L}) d\mathcal{L} = (1+z)^D n_o, \quad (3)$$

where $n_o = n(0)$. We assume that $0 < \mathcal{L}_{min} \leq \mathcal{L} \leq \mathcal{L}_{max} < \infty$.

The volume integrations are most conveniently carried out in terms of the comoving radial distance $\chi$ or the dimensionless conformal time $\eta$. The relationship between these quantities and the redshift is (e.g. Weinberg 1972)

$$\chi = \eta_o - \eta; \quad (1+z) = R_o/R(\eta), \tag{4}$$

where $\eta_o$ and $R_o = R(\eta_o)$ are the present conformal time and present expansion scale factor. The total number of sources located at comoving distances between $\chi$ and $\chi + d\chi$ is (e.g. Mészáros & Mészáros 1988) $4\pi n(\chi) R^3(\eta) f^2(\chi) d\chi$, where $f(\chi)$ is either $\sinh \chi$, $\chi$, or $\sin \chi$ for an open, flat or closed universe. Due to time dilation, the number per year of bursts at the observer position $\chi = 0$ is $(1+z)^{-1}$ times smaller (Mao & Paczyński 1992), or $(1+z)^{-1} 4\pi n(\chi) R^3(\eta) f^2(\chi) d\chi$. Hence the total number of GRB observed per year at $\chi = 0$ out to comoving distances $\leq \chi_1$ (the integrated GRB count rate) is

$$N'(<\chi_1) = 4\pi n_o R_o^3 \int_0^{\chi_1} (R(\eta)/R_o)^{4-D} f^2(\chi) d\chi. \tag{5}$$

In the Euclidean case, the bolometric peak count rate $C$ (in photons cm$^{-2}$ s$^{-1}$) observed from a GRB at distance $d$ with bolometric peak photon emission rate $\mathcal{L}$ photons per second is given by $C = \mathcal{L}/(4\pi d^2)$. The corresponding expression for the Friedmann model case for a GRB at comoving distance $\chi$ is (aside from possible K-corrections)

$$C = \frac{\mathcal{L}}{4\pi d_c^2} = \frac{\mathcal{L}}{4\pi d_p^2 (1+z)} = \frac{\mathcal{L} R(\eta)}{4\pi R_o^3 \chi^2}, \tag{6}$$

where $d_p = R_o \chi$ is the proper distance (Weinberg 1972) out to the redshift $z$ corresponding to the comoving distance $\chi$, $(1+z)$ is the redshift factor accounting for cosmological time dilation, and we have defined a photon "count distance" $d_c = d_p(1+z)^{1/2}$ related to the more usual "luminosity distance" $d_l = d_p(1+z)$ by $d_c = d_l(1+z)^{-1/2}$.

## 3. STANDARD CANDLE DISTRIBUTION IN A FLAT UNIVERSE

The simplest luminosity distribution is one where GRBs are standard candles, $\Phi(\mathcal{L}) = n_o \delta(\mathcal{L} - \mathcal{L}_o)$, where $\int_{\mathcal{L}_{min}}^{\mathcal{L}_{max}} \delta(\mathcal{L} - \mathcal{L}_o) d\mathcal{L} = 1$. For the Einstein-de Sitter flat universe, the function $f(\chi) = \chi$, and one can take $\eta_o = 1$; $R_o = 2c/H_o$; $R(\eta) = R_o \eta^2 = R_o(1-\chi)^2$, where $c$ is the velocity of light and $H_o$ is the present Hubble parameter, while the count distance is

$$d_c = d_p(1+z)^{1/2} = d_l(1+z)^{-1/2} = R_o \chi (R_o/R(\eta))^{1/2} = R_o(\sqrt{1+z} - 1). \tag{7}$$

In this special (flat universe) case, the counts received from $\chi$ are therefore $C = (\mathcal{L}_o/(4\pi R_o^2))((1-\chi)^2/\chi^2)$, and the number of bursts observed per year originating from comoving distances up to $\chi_1$ is

$$N(<\chi_1) = 4\pi n_o R_o^3 \int_0^{\chi_1} (1-\chi)^{2(4-D)} \chi^2 d\chi, \tag{8}$$

assuming that the burst spatial distribution sample is brightness limited (or rather peak count limited, $\chi_1 \ll \chi_{max}$; for a redshift limited sample, see Appendix B). Considering integer values $0 \leq D \leq 4$ and substituting $\chi_1 = [1 + R_o(4\pi C/\mathcal{L}_o)^{1/2}]^{-1}$, eq.(8) gives a closed expression for the integrated burst count rates per year

$$N(>C) = \frac{4\pi}{3} n_o \frac{\mathcal{L}_o^{3/2}}{(4\pi C)^{3/2}} (1 + (\mathcal{L}_o/(4\pi C R_o^2))^{1/2})^{-3} \left[ \sum_{k=0}^{(8-2D)} a_k \ (1 + (4\pi C R_o^2/\mathcal{L}_o)^{1/2})^{-k} \right] \ ;$$

$$a_k = \frac{(-1)^k 3}{(k+3)} \frac{(8-2D)!}{k!(8-2D-k)!}. \tag{9}$$

Equ.(9) defines the function $N(>C)$ of the independent variable $C$ with three parameters $n_o$, $\mathcal{L}_o$, and $D$. (This expression and the definition of $a_k$ is valid also for half-integer values of $D = 1/2, 3/2, 5/2, 7/2$, besides the integer values $D = 0, 1, 2, 3, 4$ used here as examples). For non-integer $D$ the integral in equ.(8) can also be calculated analytically in particular cases, and numerically in general. The analytic expression (9) with integer values of $0 \leq D \leq 4$ delimits the range of behavior expected for the integrated burst count rates over most cases of interest. For the special case $D = 3$ (constant comoving GRB density), equ.(9) reduces to a form previously derived by Mao & Paczyński (1992) and Wasserman (1992). Note that, if $\alpha \neq 1$, $D$ in equations (5) and (8) can just be replaced by $D_{eff} = (D + \alpha - 1)$, and in equ.(8) the integration can be done very simply for integer and half-integer $D_{eff}$. In this case, one needs in addition to use $\mathcal{L} = (1+z)^{\alpha-1}\mathcal{L}_K$ in eq.(6), and hence here the relation between $\chi$ and $\mathcal{L}_K = \mathcal{L}_o$ is given for any $C$ by $C = (\mathcal{L}_o/(4\pi R_o))^2((1-\chi)^{4-2\alpha}/\chi^2)$. Calculating $\chi$ from this relation, and substituting it into the integrated version of equ.(8), one obtains the analogue of equation (9). In what follows, however, we continue to use for simplicity the case $\alpha = 1$.

In the limit of very large and very small $C$ the integrated burst counts tend to

$$\lim_{C \to \infty} N(>C) = (4\pi/3) n_o \mathcal{L}_o^{3/2} (4\pi C)^{-3/2} \ ;$$
$$\lim_{C \to 0} N(>C) = (4\pi/3) n_o R_o^3 \ [3/(11-2D) \quad -6/(10-2D) + 3/(9-2D)]. \tag{10}$$

(The second limit can be obtained from equ.(8) by taking for the upper limit $\chi_1 = 1$ corresponding to $z \to \infty$; the square bracket in (10) is identical to the $A_D$ defined in eq.(21)). This means that the theoretical $\log N(>C)$ vs. $\log C$ curve is a monotonically decreasing function, with $d(\log N(>C))/d(\log C)$ monotonically decreasing from 0 to $-\frac{3}{2}$ as $C$ runs from 0 to $\infty$ (as $\log C$ runs from $-\infty$ to $\infty$). In reality, of course, $N(>C)$ as a function of $C$ is defined only for $C \geq C_{th} > 0$, where $C_{th}$ is the observational threshold peak photon count rate.

The relation (9) can be used for a limited test of the cosmological interpretation of GRB, under the assumption of a flat universe in the brightness limited standard candle case. Since $\chi_1 = 1 - (1+z)^{-1/2}$, we can rewrite equ.(9) in the form

$$N(>C) = \frac{4\pi n_o}{3} \frac{\mathcal{L}_o^{3/2}}{(4\pi C)^{3/2}} \ q(z) =$$

$$\frac{4\pi n_o}{3} \frac{\mathcal{L}_o^{3/2}}{(4\pi C)^{3/2}} \left[ (1+z)^{-3/2} \sum_{k=0}^{(8-2D)} a_k \ (1-(1+z)^{-1/2})^k \right], \tag{11}$$

where $0 \leq q(z) \leq 1$; $q(0) = 1$ and $a_k$ is defined in eq.(9). The function $q(z)$ (the square bracket in equ. 11) quantifies the degree of departure of the curve $N(>C)$ vs. $C$ from the Euclidean homogeneous unbounded case, for which $N(>C) \sim C^{-3/2}$.

It is instructive to carry out a rough visual comparison of the data against the simple theoretical model considered in this section. For any $C \geq C_{th}$ one has from the BATSE data an observed value of $N(>C)$, whereas from the extrapolated -3/2 relation one obtains for any $C \geq C_{th}$ a second "Euclidean" value $N_E(>C)$. The ratio $N(>C)/N_E(>C)$ must be given, under the standard candle assumption, by the function $q(z)$. Determining $q(z)$ from a comparison of the observed and Euclidean GRB burst rates one may determine therefore the $z$ corresponding to a given $C$. For $D=4$ the expression for $z$ from $q(z)$ is analytical, while for the remaining values of $D$ it can be obtained numerically. The values of $z$ corresponding to each $C$ thus determined from the observations should, in a true unbounded standard candle flat model, be related to the photon emission rate $\mathcal{L}_o$ through

$$\mathcal{L}_o = 4\pi R_o^2 C[(1+z)^{1/2} - 1]^2 = 4 \times 10^{57} h^{-2} C[(1+z)^{1/2} - 1]^2 \text{ photons s}^{-1}, \qquad (12)$$

where $h = H_o/(100 \text{ km s}^{-1} \text{ Mpc}^{-1})$, and where $C$ is in units of photons/(cm$^2$ s). (If the average photon has an energy of $\epsilon$ MeV, then $L_o = 6.4 \times 10^{51} C h^{-2} \epsilon [(1+z)^{1/2}-1]^2$ ergs/s is the peak bolometric luminosity in energy units). This procedure can be carried out for all $C \geq C_{th}$ and for different $D$. If the results gave a constant $\mathcal{L}_o$ (or $L_o$) for a given $D$, one would conclude that the BATSE data are compatible with the simplifying assumptions of the above model.

The results of this exercise, when applied to the publicly available BATSE (2B) data are indicated in Table 1 for some specific values of $C$ (c.f. also Figure 1). Values of $C$ above 5 photons cm$^{-2}$ s$^{-1}$ were not included, since here the departure from $C^{-3/2}$ is difficult to determine with the present approximate graphical eye-fit. From an inspection of the numbers in Table 1 it follows that there appears to be no drastic difference between the cases $D = 3$ and $D \neq 3$. A qualitatively similar conclusion has been reached, based on numerical integrations of the integrated burst counts, by Dermer (1992) and Piran (1992). One might conclude, very roughly, that $\mathcal{L}_o \sim const.$ for any $D$ when $C$ is in the interval (0.3-few) photons/(cm$^2$s). Nevertheless, one gets systematically smaller values of $\mathcal{L}_o$ for $C = (2-5)$ photons/(cm$^2$s). For these large $C$ the unsatisfactory fit of the theoretical curve of equ.(10) with the BATSE (2B) data may be caused by the assumption of a standard candle, it could be that the function $q(z)$ is systematically larger than assumed here, or it may be that for $C = (2-5)$ photons/(cm$^2$s) a better treatment of the low number statistics and errors is needed. To decide between these options would require going beyond the present analytical and qualitative considerations, e.g. a more detailed statistical fit. This is currently in preparation, and will be presented elsewhere. However, it would not be surprising if the basic premises of the model in this Section are too simple. In the next Section we consider a more complete model, by relaxing the assumption of a standard candle assumption.

## 4. THE EFFECT OF A LUMINOSITY FUNCTION

Consider now a specific example of a photon emissivity distribution similar to a Schechter function,

$$\Phi(\mathcal{L}) = \bar{n} \mathcal{L}_{min}^{-1} (\mathcal{L}/\mathcal{L}_{min})^{-\beta} , \qquad (13)$$

where $\mathcal{L}_{min} \leq \mathcal{L} \leq \mathcal{L}_{max}$, and outside of this interval $\Phi(\mathcal{L}) = 0$. We assume that $\bar{n} = const$ and $\beta$ is a dimensionless real constant. For the rest of this section we will further assume that $f(\chi) = \chi$, i.e. we take a flat universe.

Consider a given peak count $C \geq C_{th}$. Then the total number of GRB seen per year that are nearer than the comoving distance $\chi_1$ and have peak emissivities between $\mathcal{L}$ and $(\mathcal{L}+d\mathcal{L})$ is given by (see equs.(5, 7-9))

$$N'(<\chi_1, \mathcal{L}) = 4\pi \Phi(\mathcal{L}) d\mathcal{L} \int_0^{\chi_1} R_o^3 (1-\chi)^{2(4-D)} \chi^2 d\chi$$

$$= \frac{4\pi}{3} R_o^3 \Phi(\mathcal{L}) d\mathcal{L} \; \chi_1^3 \; [\; 1 + \sum_{k=1}^{8-2D} a_k \; \chi_1^k \;], \tag{14}$$

where $a_k$ is given by (9). These bursters will have count rates larger than the value $C$ corresponding to the emissivity $\mathcal{L}$ at distance $\chi_1$, related through $\chi_1 = (1 + R_o(4\pi C/\mathcal{L})^{1/2})^{-1}$. Repeating this procedure for any $\mathcal{L}_{min} \leq \mathcal{L} \leq \mathcal{L}_{max}$ for the same $C$, i.e. keeping $C$ fixed but varying $\mathcal{L}$ and hence $\chi_1$ (and assuming a brightness limited sample; for a sample with a finite $z_{max}$ see Appendix B) one obtains

$$N(>C) = \frac{4\pi \mathcal{L}_{min}^{3/2} \bar{n}}{3(4\pi C)^{3/2}} \; I = \frac{4\pi}{3} \bar{n} R_o^3 b^3 \; I, \tag{15}$$

where

$$I = \int_1^K \frac{x^{3/2-\beta} dx}{(1+bx^{1/2})^3} [1 + \sum_{k=1}^{8-2D} a_k \frac{(bx^{1/2})^k}{(1+bx^{1/2})^k}]$$

$$= \frac{2}{b^{5-2\beta}} \int_b^{bK^{1/2}} \frac{y^{4-2\beta} dy}{(1+y)^3} [1 + \sum_{k=1}^{8-2D} a_k \frac{y^k}{(1+y)^k}]; \tag{16}$$

$$x = \frac{\mathcal{L}}{\mathcal{L}_{min}}, \quad y = bx^{1/2}, \quad K = \frac{\mathcal{L}_{max}}{\mathcal{L}_{min}} \gg 1; \quad b = \frac{\mathcal{L}_{min}^{1/2}}{(R_o^2 4\pi C)^{1/2}}. \tag{17}$$

The integral $I$ is a dimensionless non-negative number, depending on $b$ (i.e. on $C$), $D$ and on $K$. Then $N(>C)$, as the function of $C$, depends on parameters $\bar{n}$, $\mathcal{L}_{min}$, $\mathcal{L}_{max}$, $\beta$ and $D$.

If $0 < b < K^{-1/2} \ll 1$, i.e. $C > C_e = \mathcal{L}_{max}/(4\pi R_o^2)$,

$$I \simeq \int_1^K x^{3/2-\beta} dx \simeq \begin{cases} (K^{5/2-\beta} - 1)/(5/2 - \beta), & \text{for } \beta \neq 5/2 \text{ ;} \\ \ln K, & \text{for } \beta = 5/2 \text{ ,} \end{cases} \tag{18}$$

since here one has always $y < 1$ with $y \ll 1$ over most of the range; hence one can approximately take $1 + y \sim 1$ and one can also take the square bracket in (16) to be unity. Then from equ.(15) it follows that

$$N(>C) \simeq \frac{4\pi}{3} \frac{\mathcal{L}_{min}^{3/2} \bar{n}}{(4\pi C)^{3/2}} \times \begin{cases} (\frac{5}{2} - \beta)^{-1}[(\mathcal{L}_{max}/\mathcal{L}_{min})^{\frac{5}{2}-\beta} - 1] \propto C^{-3/2}, & \text{for } \beta \neq \frac{5}{2}; \\ \ln(\mathcal{L}_{max}/\mathcal{L}_{min}) \propto C^{-3/2}, & \text{for } \beta = \frac{5}{2}, \end{cases} \tag{19}$$

which is the Euclidean limit.

If $K^{-1/2} \leq b < 1$, or $C_f = \mathcal{L}_{min}/(4\pi R_o^2) < C < \mathcal{L}_{max}/(4\pi R_o^2) = C_e$, one may approximately write

$$I \simeq \quad 2b^{-5+2\beta} \int_b^1 y^{4-2\beta} dy + 2A_D b^{-5+2\beta} \int_1^{bK^{1/2}} y^{1-2\beta} dy$$

$$\simeq \begin{cases} (2/(2\beta - 5)) = const., & \text{for } \beta > 5/2; \\ -\ln b^2 + (2A_D/3) \simeq const., & \text{for } \beta = 5/2; \\ b^{-5+2\beta}(2/(5-2\beta) + A_D/(\beta - 1)), & \text{for } 1 < \beta < 5/2; \\ b^{-3}((2/3) + A_D \ln(b^2 K)), & \text{for } \beta = 1; \\ b^{-3} A_D K^{1-\beta}(1-\beta)^{-1}, & \text{for } \beta < 1, \end{cases} \tag{20}$$

where

$$A_D = [1 + \sum_{k=1}^{8-2D} a_k] = 3\int_0^1 (1-\chi)^{8-2D}\chi^2 d\chi$$
$$= [3/(9-2D) - 6/(10-2D) + 3/(11-2D)], \quad (21)$$

with $a_k$ given in (9). (Some particular values of $A_D$ are $A_4 = 1$, $A_3 = 1/10$, $A_2 = 1/35$, etc. For most of the range $y \leq 1$ one may use $(1+y) \simeq 1$, while for $y \geq 1$ one can take $(1+y) \simeq y$. For $\beta > (5/2)$ the second integral is much smaller than the first one, because $A_D \leq 1$ and $b < 1$. For $\beta < 1$ the first integral is much smaller than the second one, because $b^2 K > 1$. Roughly one may also take $\ln b \simeq const$. In fact, one can check the validity of this approximation for several $\beta$ by direct integration. For example, for $\beta = (3/2)$ and $D = 3$ one has exactly for $b \ll 1$ and $bK^{1/2} \gg 1$ $I = 0.72b^{-2}$. Equ.(20) gives $I = 1.2b^{-2}$.) Substituting the value of $I$ from equ.(20) into equ.(15) one obtains

$$N(>C) \simeq (4\pi\bar{n}/3) \times$$

$$\begin{cases} \mathcal{L}_{min}^{3/2}(\beta - 5/2)^{-1}(4\pi C)^{-3/2} \propto C^{-3/2}, & \text{for } \beta > 5/2; \\ \mathcal{L}_{min}^{3/2}(4\pi C)^{-3/2}[(2A_D/3) - \ln(\mathcal{L}_{min}/(R_o^2 4\pi C))] \propto C^{-3/2}, & \text{for } \beta = 5/2; \\ \mathcal{L}_{min}^{\beta-1} R_o^{5-2\beta}(4\pi C)^{1-\beta}[(5/2 - \beta)^{-1} + (A_D/(\beta - 1))] \propto C^{1-\beta}, & \text{for } 1 < \beta < 5/2; \\ R_o^3[A_D \ln(\mathcal{L}_{max}/(4\pi C R_o^2)) + (2/3)] \simeq const, & \text{for } \beta = 1; \\ R_o^3 A_D(1-\beta)^{-1}(\mathcal{L}_{max}/\mathcal{L}_{min})^{1-\beta} = const, & \text{for } \beta < 1. \end{cases} \quad (22)$$

A similar behavior for the range $1 \leq \beta \leq 5/2$ had been previously derived for a specific value of $D = 3$ (Wasserman 1992), but here we see from equ.(22) that the scaling $N(>C) \propto C^{1-\beta}$ is independent of $D$, i.e. it is valid for an arbitrary redshift dependence of the comoving density.

For $b \gg 1$, or $C \ll C_f = \mathcal{L}_{min}/(4\pi R_o^2)$ one may write

$$I \simeq A_D b^{-3} \int_1^K x^{-\beta} dx, \quad (23)$$

where $A_D$ is given in eq.(21). Hence

$$N(>C) \simeq (4\pi/3)\bar{n}R_o^3 A_D \times \begin{cases} (\beta - 1)^{-1}[1 - (\mathcal{L}_{max}/\mathcal{L}_{min})^{1-\beta}] = const. & \text{for } \beta \neq 1; \\ \ln(\mathcal{L}_{max}/\mathcal{L}_{min}) = const. & \text{for } \beta = 1. \end{cases} \quad (24)$$

Equs.(19), (22) and (24) define the integral GRB count rate function $N(>C)$ of the independent variable $C$; this function depends on the parameters $\bar{n}$, $D$, $\mathcal{L}_{max}$, $\mathcal{L}_{min}$ and $\beta$.

## 5. DISCUSSION

The purpose of this paper is to provide analytical expressions characterizing the various regimes that would be present in a cosmological distribution of bursting sources with some generic luminosity and density distributions. With a simple redefinition of the density parameter $D$ in eq.(2) to $D_{ss} = D + 1$, the expressions are also valid for steady (non-bursting) sources. The specific example discussed is that of a cosmological distribution of gamma-ray burst sources. We do not present detailed statistical fits here, concentrating instead on extracting insights and quantitative estimates from the analytical models developed above, and analyze the effect of the physical assumptions and the parameters on the various asymptotic behaviors expected for the brightness distribution (log N- log C) of the bursts.

The simplest standard candle model is given here a general analytical representation, which complements previous analyses by providing a specific dependence of the counts on the density evolution with redshift, as well as on the cosmological curvature (see App. A). This expression quantifies specifically the departures from a simple homogeneous Euclidean model as a function of either redshift or count rate. The $\log N(>C) - \log C$ curve has two asymptotic behaviors, the Euclidean $C^{-3/2}$ for high $C$ (nearby sources) and $C^0$ for very low $C$ (distant sources) and $D < 9/2$. The flattening, caused by cosmological redshift effects, is very gradual, being given by the function $q(z)$ (the square bracket in equ.(11)). Approximate fits to the BATSE and PVO data are possible (c.f. also Mao and Paczyński, 1992, Dermer, 1992, Piran, 1992, Fenimore & Bloom, 1995), which are not strongly constrained by the data. The asymptotic value of $N(>C)$ is given by equ.(10), and is a function of $D$, the parameter characterizing the density dependence of sources via eq.(2). This assumes brightness-limited sources, or $\chi_1(C_{th}) = (1 + [4\pi R_o^2 C_{th}/\mathcal{L}_o]^{1/2})^{-1} < \chi_{max} = (1 - (1+z_{max})^{-1/2})$, where $z_{max}$ is the redshift at which the most distant (but unobservable) sources are. Otherwise, for a redshift-limited sample, the $N(>C)$ also becomes constant at the $C$ below those corresponding to the weakest sources at $z_{max}$, but the limiting value of $N(>C)$ is smaller than (10) (see Appendix B). One advantage of the present analytical treatment is that, given an ideal set of data and assuming the brightness-limited standard candle case of §3, a simple and direct estimate of a lower limit to $z_{max}$ could be obtained from eq.(11) as a function of the density evolution exponents $D$. One would do this by finding the $z$ values at which flattening starts to set in (see also Table 1). In practice, however (see Fig. 1), this flattening regime is not clearly seen at the lowest $C$ considered, and the data are not "ideal". While the standard candle model gives reasonable numbers ($L_o \sim 6 \times 10^{51}$ erg/s, $z_{max} \gtrsim 1$, depending on $D$), the present approximate fit is not unique (see §3). A detailed statistical fit is needed in order to get more quantitative constraints on such models (e.g. Emslie & Horack 1994, Cohen & Piran, 1995). However, a number of instrumental uncertainties and incompleteness consideration must be taken into account, especially at low $C$, and the constraints so far are weak.

In the presence of a power-law luminosity function (13), equations (18-24) show that the relation $\log N(>C)$ vs. $\log C$ can have three different asymptotic behaviors, instead of two. A simple physical interpretation of this behavior for homogeneous Euclidean space is discussed in Wasserman (1992), which also indicates what can be expected in cosmology. Here, we have obtained the analytic behavior for cosmologies with a density evolution that can depart from a constant comoving density ($D \neq 3$ as well as $D = 3$), and have evaluated these affects also for $\Omega_o \leq 1$. The values of the constants in the expressions for the counts (eqs. (19), (22), (24)) depend on $D$ for a brightness limited sample. For a redshift limited sample the constants are somewhat different and can depend explicitly on $z_{max}$ (e.g. Appendix B), and for non-flat cosmologies they can also depend on $\Omega_o$ (see App. A). The slopes of $(\log N(>C))$ increase from $-3/2$ at high $C$ towards 0 at low $C$, as in the standard candle case, but now for a luminosity function index $1 < \beta < (5/2)$ the integral counts go through an intermediate linear behavior where the slope is $(1 - \beta)$, before flattening to 0 (for $D < 9/2$) at very low $C$, i.e., the behavior can be approximated by three straight lines rather than two.

The first transition, from the Euclidean $C^{-3/2}$ behavior to the $C^{1-\beta}$ behavior, occurs at

$$C_e \simeq \frac{\mathcal{L}_{max}}{4\pi R_o^2} Q_e, \tag{25}$$

where $Q_e \simeq \left[1 + A_D((5/2) - \beta)/(\beta - 1)\right]^{1/(\beta - (5/2))} \simeq 1$. (For $\beta \geq (5/2)$ the behavior $C^{-3/2}$ continues until the next transition $C_f$ below; thus here no analogy of $C_e$ exists. For $\beta < 1$ equation (25) may also be used to define $C_e$ with $Q_e \simeq (A_D((5/2) - \beta)(1 - \beta)^{-1})^{-2/3}$; here $C_e$ defines the break between the $C^{-3/2}$ and $C^0$ behavior.). The observed $\log N - \log C$ curve (Fig. 1) can be used for deriving an approximate estimate of the first break, $C_e \sim 6.9$ photons cm$^{-2}$ s$^{-1}$, which from eq.(25) yields

$$\mathcal{L}_{max} \simeq 2.5 \times 10^{58} h^{-2} (C_e/6.9) \; s^{-1}. \tag{26}$$

This corresponds, assuming a typical photon energy $\epsilon$ MeV, to a luminosity $L_{max} \simeq 2.5 \times 10^{52} \epsilon h^{-2}(C_e/6.9)$ erg s$^{-1}$ (c.f. also Wasserman, 1992).

The second transition from the $C^{1-\beta}$ behavior to the flat asymptotic behavior $C^0$ occurs at

$$C_f \simeq \frac{\mathcal{L}_{min}}{4\pi R_o^2} Q_f, \qquad (27)$$

where $Q_f \simeq \left[1 + (\beta - 1)((5/2) - \beta)^{-1} A_D^{-1}\right]^{1/(\beta-1)}$ for $1 < \beta < 5/2$. (Equ.(27) may be defined also for $\beta > 5/2$, with $Q_f \simeq \left[(\beta - 1)/([\beta - 5/2]A_D)\right]^{2/3}$. In this case $C_f$ defines the break between the $C^{-3/2}$ and $C^0$ behavior. No analogy of $C_f$ exists for $\beta \leq 1$.) This second break point cannot be clearly defined from the data in Fig. 1, but an upper limit to it may be taken to be $C_f \lesssim 0.4$ photons cm$^{-2}$ s$^{-1}$. (We note that the 2B catalogue has some data points at $C \lesssim 0.4$ photons cm$^{-2}$ s$^{-1}$, for which however the triggering corrections become very uncertain, so we did not use any data points below this value).

The specific value of $\beta$ is most important in the range $1 < \beta < 5/2$ because in this case there can be a range of $C$ intermediate between the regimes $C^{-3/2}$ and $C^0$ where the behavior of $N(>C)$ reflects directly the effects of the luminosity function sampled at the largest redshifts. (Note that even for $z_{max} \to \infty$ as in §4 the relevant largest redshifts are $z \sim 3$ because that is where $C_e \simeq \mathcal{L}_{max}/(4\pi R_o^2)$ is equivalent to $C(\mathcal{L}_{max}) = \mathcal{L}_{max}/(4\pi d_c^2(z_{max})) = \mathcal{L}_{max}/(4\pi R_o^2[\sqrt{1+z} - 1]^2))$. On the other hand, for $\beta \geq 5/2$ the $N(>C)$ curve is dominated by the faintest bursts, while for $\beta \geq 1$ it is dominated by the brightest bursts in the luminosity function. From the data in Fig. 1, one can obtain an approximate fit to the intermediate slope of $(1 - \beta) \sim -0.88$, or $\beta \sim 1.88$ (c.f. also Meegan et al. 1992, Wasserman, 1992). Note, however, that the credible contours for parameters of a wide variety of models are so far extremely large (e.g. Emslie & Horack, 1994, Cohen & Piran, 1995), so there is a degeneracy of possible models. For instance, Loredo and Wasserman (1992) make the point, also apparent from our analysis, that it is difficult to distinguish between a standard candle with no evolution and ones with a luminosity function, or standard candles with density evolution. A recent discussion of some of the relevant issues of data analysis methodology and incompleteness problems has been given by Loredo and Wasserman (1995).

If one uses a fit with a power-law slope of $1 - \beta \sim -0.88$, and assuming for the sake of argument $D = 3$ so that $A_3 = 10^{-1}$ and $Q_f \sim 2.2 \times 10^1$, an upper limit for $\mathcal{L}_{min}$ is

$$\mathcal{L}_{min} \lesssim 7 \times 10^{55} h^{-2}(C_f/0.4)(Q_f/22)^{-1} \, s^{-1}, \qquad (28)$$

corresponding to a luminosity $L_{min} \lesssim 7 \times 10^{49} \epsilon h^{-2}(C_f/0.4)(Q_f/22)^{-1}$ erg s$^{-1}$. The corresponding lower limit on the ratio of the maximum to minimum photon emissivity in the distribution function (13) is

$$K \equiv (\mathcal{L}_{max}/\mathcal{L}_{min}) \gtrsim 3 \times 10^2 (C_e/6.9)(C_f/0.4)^{-1}(Q_f/22) \,. \qquad (29)$$

From Figure 1 we also obtain a lower limit value of $N(>C_f) \gtrsim 3 \times 10^2$ averaged over approximately two years which, from eq.(23) with $D = 3$, $A_D = 0.1$ yields a lower limit to the total burst rate density

$$\bar{n} \gtrsim 1.5 \times (A_D/0.1)^{-1} h^3 Gpc^{-3} yr^{-1} \,. \qquad (30)$$

If we assume $L_*$ galaxies to have a density of 0.01 $h^3$ Mpc$^{-3}$, this gives a lower limit to the total burst rate of $\mathcal{R} \gtrsim 1.5 \times 10^{-7}(A_D/0.1)^{-1}$ bursts per year per $L_*$ galaxy. (Note that, from eqs.(3) and (13), $\bar{n} \simeq n_o$). This lower limit can be checked also using equ.(19) at $C_e$, which gives the same limit $\mathcal{R} \gtrsim 1.5 \times 10^{-7}(K/300)^{\beta-1}$ per $L_*$ galaxy per year. The $K^{\beta-1}$ dependence arises because of the assumed $C^{1-\beta}$ dependence with $C \propto \mathcal{L}$ in this regime. The fact that these two independent estimates are similar, based on (23) evaluated for $A_D = 0.1$ and (19) which is independent of $A_D$, suggests that $D = 3$ is not an inconsistent assumption.

The same analysis of §3-4 can be done for $\Omega_o \neq 1$ (see Appendix A). The result is that small higher order terms appear in the integrals (8) and (14) which in (11) and (16) lead to additional small, $\Omega_o$ dependent correction terms in the square bracket, in addition to the previous small evolution and luminosity function correction terms, the leading term still being unity as before. These extra terms explicitly quantify the weakness of the dependence on $\Omega_o$, and explain why numerical statistical fits (e.g. Loredo & Wasserman, 1992, Emslie & Horack, 1995, Cohen & Piran, 1995) have not so far been able to constrain cosmological parameters. The reason for this insensitivity to the closure parameter is that the density, luminosity and cosmological effects are first order corrections in the comoving maximum distance $\chi_1$ (with $\chi_1 \ll 1$ (App. A) for any phenomena dependent on star formation at $z_{max} \lesssim$ few). For different cosmological models the shape of the curves remains similar, and only the absolute length scales change, i.e. the way $\chi$ depends on $z$. The latter could affect the observationally determined value of $\mathcal{L}_{max}, \mathcal{L}_{min}$ (which also depend on $h = (H_o/100)$, the value of the present Hubble constant in units of 100 Km/s/Mpc, and thus on $\Omega_o$), but it does not allow one to discriminate between distance effects and possible intrinsic physical effects on $\mathcal{L}$ at the source.

Another simplification made in §3 and §4 is to take the sources to be brightness-limited, with the density dependence $n(z) \propto (1+z)^D$ out to the largest $z$ or smallest $C$ observed. Realistically, however, one could expect a different functional dependence, or alternatively it might be that $n(z) \propto (1+z)^D$ only for $z \leq z_{max}$, while for $z > z_{max}$ one might have $n(z) \sim 0$. The straightforward generalization of the mathematical formalism of this paper to such redshift-limited cases is carried out in Appendix B. This shows that, while for $z_{max} \lesssim 3$ the values of $\bar{n}$ would be somewhat affected, most effects associated with a finite $z_{max}$ are generally small.

The luminosities and burst rate densities derived from this analysis are compatible with those expected for cosmological models based either on compact stellar merger or collapse scenarios (e.g. Narayan, Paczyński & Piran 1992, Woosley 1993). Definitive conclusions regarding this would require further detailed statistical fits, and the resolution of a number of uncertainties about the completeness and the exposure function at low fluences. Models with a broad range of $\Omega_o$, density evolution and luminosity function parameters appear acceptable. In particular, the present evidence does not appear sufficient to discriminate between a standard candle distribution (which would imply a well defined density) and a power law (or approximately similar) luminosity distribution (where the total density would be harder to estimate due to fading of counts at the low fluence end). However, some constraints can be made for specific cases. Thus, if the low fluence end of the counts is determined by a luminosity function of index $\beta \sim 1.88$ one has approximately 10% fewer sources in each decade of luminosity than in the previous lower decade, i.e. if $K \sim 10^2$ one has 90% of the sources between $\mathcal{L}_{min}$ and $10\mathcal{L}_{min}$ and 10% between $10\mathcal{L}_{min}$ and $10^2\mathcal{L}_{min} \sim \mathcal{L}_{max}$, a statement compatible with the findings of Emslie & Horack 1994 and Ulmer & Wijers 1994. We note, however, that $\mathcal{L}_{min}$ (eq.[28]) is a current upper limit, which could become lower with more sensitive observations. While this cannot be ruled out energetically, if the integrated burst rates continued with a slope $1 - \beta \sim -0.88$ down to count values much less than one order of magnitude below the present estimated upper limit $C_f \sim 0.4$ cm$^{-2}$s$^{-1}$, the required burst rate would exceed the maximum value $\mathcal{R} \sim 10^{-6} - 10^{-5}$ per year per galaxy expected from compact mergers or collapse, providing constraints on such scenarios.

**ACKNOWLEDGMENTS**

This research has been partially supported through NASA NAGW-1522 and NAG5-2362. We are grateful to M.J. Rees for useful advise and discussions. A.M. is grateful for the hospitality of the Astronomy and Astrophysics Department of Pennsylvania State University.

## APPENDIX A: EFFECTS OF A NON-FLAT COSMOLOGY

For a general Friedmann model the integrated number of bursts out to $\chi_1$ is still given by eq.(5), but $f(\chi)$ is $\sinh\chi$ ($\sin\chi$) for an open (closed) universe. The other difference is that the expansion factor (which was $R(\eta)/R_o = \eta^2$ for the flat universe) is now, assuming $\Omega_o \leq 1$,

$$\frac{R(\eta)}{R_o} = \frac{(\cosh\eta - 1)}{(\cosh\eta_o - 1)} = \frac{1}{1+z} \tag{31}$$

where $\cosh\eta_o = (2/\Omega_o) - 1$, $\chi = \eta_o - \eta$ and $\cosh\eta = \cosh(\eta_o - \chi)$. For small $\Omega_o$ we can expand

$$\frac{\cosh\eta_o - 1}{\cosh\eta - 1} \simeq 1 - g\chi, \tag{32}$$

where $g \simeq [1 + \frac{1}{2}\Omega_o] \sim 1$. Also, since for $z_1 \lesssim$ few one has $\chi_1 \ll 1$, we can take

$$f^2(\chi) = \sinh^2\chi \sim \chi^2. \tag{33}$$

With this we have the integrated number of bursts per year out to $\chi_1$ as

$$\begin{aligned} N'(<\chi_1) &\simeq 4\pi\Phi(\mathcal{L})d\mathcal{L}R_o^3 \int_0^{\chi_1}(1-g\chi)^{2(4-D)}\chi^2 d\chi \\ &\simeq \tfrac{4\pi}{3}R_o^3 \Phi(\mathcal{L})d\mathcal{L}\,\chi_1^3\left[1 + \sum_{k=1}^{2(4-D)} a_k\,(g\chi_1)^k\right], \end{aligned} \tag{34}$$

where $a_k$ is given by (9). Since $\chi_1 \ll 1$ and $g \simeq 1$, the square bracket is again $[\cdots] \sim 1$ and we get the same result as eq.(14) (to within small correction terms of order $g^{-1} \simeq 1 - (\Omega_o/2) \to 1$ for $\Omega_o \ll 1$).

Therefore $N(< C)$ is also given by eq.(16) to within the same small correction factor, and (19), (21) and (24) are also approximately correct for $\Omega_o \ll 1$. The specific dependence of $\chi$ on $z$ is, however, a function of $\Omega_o$, which affects the length and luminosity scales, but not the shapes of the curves. Other effects associated with a possible cosmological constant can also be included in a straightforward manner (e.g. Piran 1992, Cohen, Kolatt & Piran 1994, Emslie & Horack 1994), and have been ignored here for simplicity.

### APPENDIX B: THE EFFECT OF A LIMITING REDSHIFT

Consider the case when the density dependence of equ.(3) is valid only up to some maximum redshift $z_{max}$,

$$
\begin{aligned}
n(z) &= n_o(1+z)^D(1-\Theta(z/z_{max})) \ , \\
n_o &= \int_{\mathcal{L}_{min}}^{\mathcal{L}_{max}} \Phi(\mathcal{L})d\mathcal{L} = \bar{n}\int_1^K x^{-\beta}dx = \bar{n}\begin{cases} (1-\beta)^{-1}[(\mathcal{L}_{max}/\mathcal{L}_{min})^{1-\beta}-1], & \text{for } \beta \neq 1, \\ \ln(\mathcal{L}_{max}/\mathcal{L}_{min}), & \text{for } \beta = 1, \end{cases}
\end{aligned}
\tag{35}
$$

where $\Theta(z/z_{max})$ is the Heaviside function. There is in this case a limiting comoving radial coordinate $\chi_{max}$ defined unambiguously by $z_{max}$. (For the flat universe $f(\chi) = \chi$ one has $\chi_{max} = (1-[1+z_{max}]^{-1/2})$, and in what follows we restrict ourselves to this case only.) Equ.(5) is correct only for $\chi_1 \leq \chi_{max}$. For $\chi_1 > \chi_{max}$ one has to take as upper limit of integration in equ.(5) $\chi_{max}$ instead of $\chi_1$. Equs.(6-7) are again correct, but, of course, are defined only for $z \leq z_{max}$.

The standard candle model considerations of §3 can be generalized as follows. Equs.(8-9) are fulfilled only in the case when $\chi_1 \leq \chi_{max}$, corresponding to $C \geq C_F$. The value $C_F$ is defined by $\chi_{max} = (1+R_o[4\pi C_F/\mathcal{L}_o]^{1/2})^{-1}$. For $\chi_1 > \chi_{max}$, in equ.(8) the upper limit of integration is $\chi_{max}$. This means that for $C \geq C_F$, equ.(9) describes as before the function $N(> C)$, but for $C \leq C_F$ one has $N(> C) = N(> C_F) = const.$ (there is a horizontal straight line for $C < C_F$). This also means that in equ.(10) the appropriate limit for $C \to 0$ is smaller than the value given, and in equ.(11) the function $q(z)$ is also defined only for $z \leq z_{max}$.

The generalization of the power-law luminosity function model of §4 may be done as follows. Here again equ.(14) holds only for $\chi_1 \leq \chi_{max}$. When $\chi_1 > \chi_{max} = (1-(1+z_{max})^{-1/2})$, the upper limit of integration is $\chi_{max}$. For an arbitrary $C$ in the interval $0 < C < \infty$ one can define a peak photon emissivity $\tilde{\mathcal{L}}(C)$ and a critical count rate $C_E$ through

$$
\begin{aligned}
\tilde{\mathcal{L}}(C) &= C4\pi R_o^2 \chi_{max}^2(1-\chi_{max})^{-2} = C4\pi R_o^2(\sqrt{1+z_{max}}-1)^2 = C4\pi d_c^2(z_{max}), \\
C_E &= \mathcal{L}_{max}/(4\pi d_c^2(z_{max})) \ ,
\end{aligned}
\tag{36}
$$

where $d_c(z_{max}) = R_o(\sqrt{1+z_{max}}-1)$ is the count distance of eq. (7), $C_E$ is the finite z equivalent of $C_e$ and $0 < \tilde{\mathcal{L}}(C) < \infty$.

Consider first the case when $\tilde{\mathcal{L}}(C) \geq \mathcal{L}_{max}$, i.e. $C \geq C_E$ (or $b \leq [\sqrt{1+z_{max}}-1]K^{-1/2}$). This means that for any $\mathcal{L}$, where $\mathcal{L}_{min} \leq \mathcal{L} \leq \mathcal{L}_{max} \leq \tilde{\mathcal{L}}(C)$, the corresponding comoving coordinate $\chi_1$ defined by $\chi_1 = (1+R_o(4\pi C/\mathcal{L})^{1/2})^{-1}$ is smaller than $\chi_{max}$, and therefore in this case the considerations of §4 are correct without any change. In other words, for $C \geq C_E$ the function $N(> C)$ vs. $C$ is unchanged. (Some numerical examples are useful here. For example, for $\chi_{max} = 0.9$ ($z_{max} = 99$), one has $C_E = (1/81)\mathcal{L}_{max}/(4\pi R_o^2)$. Then, if $K \leq 81$, equs.(18-21) remain unchanged. But, if $\chi_{max} = 0.5$ ($z_{max} = 3$), one has $C_E = \mathcal{L}_{max}/(4\pi R_o^2) \equiv C_e$ ss defined in equ.(25), and only the Euclidean part of equs.(18-19) remain unchanged. For $z_{max} < 3$ even the Euclidean part is influenced by the existence of $z_{max}$. Nevertheless, the regimes $C^{-3/2}$, $C^{1-\beta}$ and $C^0$ still occur, but in modified form and with different normalizations and break points; see below).

Next, take $\mathcal{L}_{min} \leq \tilde{\mathcal{L}}(C) \leq \mathcal{L}_{max}$, or $C_F = \mathcal{L}_{min}/(4\pi d_c^2(z_{max})) \leq C \leq \mathcal{L}_{max}/(4\pi d_c^2(z_{max})) = C_E$ (or, equivalently, $[\sqrt{1+z_{max}} - 1]K^{-1/2} \leq b \leq [\sqrt{1+z_{max}} - 1]$). For $\mathcal{L}_{min} \leq \mathcal{L} \leq \tilde{\mathcal{L}}(C)$ equ.(14) still holds, because here $\chi_1 \leq \chi_{max}$. On the other hand, for $\tilde{\mathcal{L}}(C) \leq \mathcal{L} \leq \mathcal{L}_{max}$ in equ.(14) the upper limit of integration is $\chi_{max}$. Hence, using the distribution (13), it follows that

$$N(>C) = \frac{4\pi \bar{n} \mathcal{L}_{min}^{3/2}}{3(4\pi C)^{3/2}} \int_1^{\tilde{\mathcal{L}}(C)/\mathcal{L}_{min}} \frac{x^{\frac{3}{2}-\beta}dx}{(1+bx^{1/2})^3} \left[1 + \sum_{k=1}^{8-2D} a_k \frac{(bx^{1/2})^k}{(1+bx^{1/2})^k}\right]$$

$$+ \frac{4\pi}{3} R_o^3 \bar{n} \chi_{max}^3 \left[1 + \sum_{k=1}^{8-2D} a_k \; \chi_{max}^k \right] \int_{\tilde{\mathcal{L}}(C)/\mathcal{L}_{min}}^K x^{-\beta}dx$$

$$= \frac{4\pi \mathcal{L}_{min}^{3/2} \bar{n}}{3(4\pi C)^{3/2}} I_1 + \frac{4\pi}{3} R_o^3 \bar{n} \left(1 - \frac{1}{\sqrt{1+z_{max}}}\right)^3 \left[1 + \sum_{k=1}^{8-2D} a_k \left(1 - \frac{1}{\sqrt{1+z_{max}}}\right)^k \right] I_2$$

$$= \frac{4\pi}{3} R_o^3 \bar{n} \; b^3 \; I_1 + \frac{4\pi}{3} R_o^3 \bar{n} \; B(z_{max}, D) \; I_2, \qquad (37)$$

where $b$ is defined by equ.(17), $\tilde{\mathcal{L}}(C)$ is defined by equ.(36), and $B(z_{max}, D)$ is a constant. We see that $N(>C)$ is defined by the sum of two terms having different behaviors. The first term is practically identical to the relation defined by equs.(15-16); the only difference follows from the upper limit of integration in $I_1$. The second term is exactly calculable; the dependence on $C$ follows from the fact that the lower limit of integration in $I_2$ depends on $C$ via $\tilde{\mathcal{L}}(C)$. The integral $I_2$ is given by

$$I_2 = \int_{\tilde{K}}^K x^{-\beta}dx = \begin{cases} (1-\beta)^{-1}[(\mathcal{L}_{max}/\mathcal{L}_{min})^{1-\beta} - (\tilde{\mathcal{L}}(C)/\mathcal{L}_{min})^{1-\beta}], & \text{for } \beta \neq 1, \\ \ln(\mathcal{L}_{max}/\tilde{\mathcal{L}}(C)), & \text{for } \beta = 1, \end{cases} \qquad (38)$$

where, for brevity, we have denoted $\tilde{K} \equiv (\tilde{\mathcal{L}}(C)/\mathcal{L}_{min})$. The integral $I_1$ may be calculated as in §4. There are three different cases. First, if $z_{max} < 3$ (or $\chi_{max} < 0.5$), then $b < 1$, and, in addition, $bx^{1/2} = (\chi_{max}/(1-\chi_{max})) = (\sqrt{1+z_{max}} - 1) < 1$ for $x = \tilde{\mathcal{L}}(C)/\mathcal{L}_{min}$. In this case one may write

$$I_1 \simeq \begin{cases} (5/2-\beta)^{-1}[\tilde{\mathcal{L}}(C)/\mathcal{L}_{min})^{5/2-\beta} - 1] & \text{for } \beta \neq 5/2, \\ \ln(\tilde{\mathcal{L}}(C)/\mathcal{L}_{min}), & \text{for } \beta = 5/2. \end{cases} \qquad (39)$$

Second, consider the range $3 \leq z_{max} \leq ((1+K^{1/2})^2 - 1)$ (or $0.5 \leq \chi_{max} \leq K^{1/2}/(1+K^{1/2})$). In this case both $b < 1$ and $b \geq 1$ may occur; but always $bx^{1/2} = \chi_{max}/(1-\chi_{max}) \geq 1$ for $x = \tilde{\mathcal{L}}(C)/\mathcal{L}_{min}$. Therefore, if $b < 1$, one has

$$I_1 \simeq \begin{cases} (2/(2\beta-5)) = const., & \text{for } \beta > 5/2; \\ -\ln b^2 + (2A_D/3) \simeq const., & \text{for } \beta = 5/2; \\ b^{-5+2\beta}(2/(5-2\beta) + A_D/(\beta-1)), & \text{for } 1 < \beta < 5/2; \\ b^{-3}((2/3) + A_D \ln(b^2 \tilde{K})), & \text{for } \beta = 1; \\ b^{-3} A_D \tilde{K}^{1-\beta}(1-\beta)^{-1}, & \text{for } \beta < 1, \end{cases} \qquad (40)$$

where we have proceeded as in the derivation of (20). The results are also the same, except for the fact that $K$ is substituted by $\tilde{K}$ for $\beta \leq 1$. In the same range of $z_{max}$ but for $b \geq 1$, instead of eq.(40) one obtains

$$I_1 \simeq A_D b^{-3} \int_1^{\tilde{K}} x^{-\beta} dx \simeq b^{-3} \times \begin{cases} (\beta - 1)^{-1}[1 - (\tilde{\mathcal{L}}/\mathcal{L}_{min})^{1-\beta}], & \text{for } \beta \neq 1, \\ \ln(\tilde{\mathcal{L}}/\mathcal{L}_{min}), & \text{for } \beta = 1. \end{cases} \qquad (41)$$

Third, consider $z_{max} > ((1 + K^{1/2})^2 - 1)$ (or $\chi_{max} > K^{1/2}/(1 + K^{1/2})$). In this case one necessarily has $b > 1$, and $I_1$ is again given by equ.(41). Substituting $I_1$ and $I_2$ into equ.(37) one obtains the function $N(>C)$ vs. $C$, which depends now on the parameters $\mathcal{L}_{min}$, $\mathcal{L}_{max}$, $\beta$, $\bar{n}$, $z_{max}$, and $D$ in analytical form within this range of $C$.

Finally, consider the case when $\tilde{\mathcal{L}}(C) \leq \mathcal{L}_{min}$; i.e. $C \leq C_F = \mathcal{L}_{min}/(4\pi d_c^2(z_{max}))$, with $d_c$ given by eq.(7), and $C_F$ is the finite z equivalent of $C_f$. In this case, one always has $\chi_1 \geq \chi_{max}$. Therefore, here equ.(14) does not hold, because the upper limit of integration is $\chi_{max}$. Hence one obtains

$$N(>C) = \frac{4\pi}{3} R_o^3 \bar{n} \chi_{max}^3 \Big[1 + \sum_{k=1}^{8-2D} a_k\, \chi_{max}^k\, \Big] \int_1^K x^{-\beta} dx = \frac{4\pi}{3} R_o^3 n_o B(z_{max}, D) = const, \qquad (42)$$

where $n_o$ is given by equ.(35) and $B(z_{max}, D)$ by equ.(37). This means that here, as in the standard candle case, there is again a horizontal straight line on the graph $N(>C)$ vs. $C$. (The differential counts $N(C)$ at $C < C_F$ are expected to be zero, while the integral counts $N(>C)$ of (42) are constant and given by the value at $C = C_F$).

These relations would allow one in principle to determine the parameters $\mathcal{L}_{min}$, $\mathcal{L}_{max}$, $\beta$, $D$, $\bar{n}$ and $z_{max}$. For instance, $\beta$ could be determined similarly to the case with $z_{max} = \infty$ with large $C$ from the asymptote of the slope (if $1 < \beta < (5/2)$). Then from the transition in the graph $N(>C)$ vs. $C$ between the $C^{-3/2}$ to the $C^{1-\beta}$ behavior one can determine $\mathcal{L}_{max}$ (if $z_{max} \geq 3$ equ.(26) holds). Using this, from the Euclidean part, one obtains $\bar{n}$ (see equs.(15) and (18)). In the ideal case then $\chi_{max}$ (or $z_{max}$) could be obtained from a careful analysis of the departure of the $N(>C)$ graph from a straight line with slope $(1 - \beta)$, because this departure should be given by the second term in equ.(37), which depends on $\chi_{max}$. Then the horizontal $C^0$ straight line at very small $C$, and its break to the $C^{1-\beta}$ behavior, could yield the value of $K$ (hence of $\mathcal{L}_{min}$) and $D$ via equs.(37), (38) and (42).

In practice, there would be severe difficulties due to incompleteness and errors at the lowest count rates. It may be very hard, if not impossible, to determine the $C^0$ straight line expected at the lowest $C$. Even in such a less than ideal $N(>C)$ graph, the formulas of this Appendix would provide a strategy for obtaining bounds on $z_{max}$. To illustrate this, consider the case when $1 < \beta < 5/2$, and $z_{max} \leq 3$. In this case the function $N(>C)$ is given by

$$N(>C) \simeq (4\pi/3) \times$$

$$\begin{cases} \frac{n_o \mathcal{L}_{min}^{3/2}}{(4\pi C)^{3/2}} \propto C^{-3/2}, & \text{for } C \geq C_E, \\ [\frac{\bar{n}\mathcal{L}_{min}^{3/2}}{(4\pi C)^{3/2}} \frac{[(\tilde{\mathcal{L}}(C)/\mathcal{L}_{min})^{5/2-\beta} - 1]}{5/2 - \beta} \\ + R_o^3 \bar{n} B(z_{max}, D) \frac{[(\mathcal{L}_{max}/\mathcal{L}_{min})^{1-\beta} - (\tilde{\mathcal{L}}(C)/\mathcal{L}_{min})^{1-\beta}]}{1 - \beta}], & \text{for } C_F \leq C \leq C_E, \\ R_o^3 n_o B(z_{max}, D) = const, & \text{for } C \leq C_F, \end{cases} \qquad (43)$$

where $C_E = \mathcal{L}_{max}/(4\pi d_c^2(z_{max}))$, $C_F = \mathcal{L}_{min}/(4\pi d_c^2(z_{max}))$ are the equivalents of $C_e$, $C_f$ in §4, $d_c(z_{max}) = R_o(\sqrt{1 + z_{max}} - 1)$, $\tilde{\mathcal{L}}(C)$ is defined by equ.(36), and $n_o = \bar{n}(1 - \beta)^{-1}[K^{1-\beta} - 1]$. We see that the Euclidean part is again continued by an approximate straight line with slope $(1 - \beta)$ for $C_F \lesssim C \lesssim C_E$.

However, due to the second term in the square bracket of equ.(42), there is a departure from this straight slope $(1-\beta)$ as $C$ decreases. If the data allowed it, a careful analysis might determine this second term from this departure, and hence also $z_{max}$. If, in addition, a $C^0$ portion were unequivocally discernible (which may not be the case), an additional relationship would be obtained linking $z_{max}$, $D$ and $n_o$.

## Table 1

| $C$ | 5.0 | 4.0 | 3.0 | 2.0 | 1.5 | 1.0 | 0.8 | 0.6 | 0.5 | 0.4 | 0.3 |
|---|---|---|---|---|---|---|---|---|---|---|---|
| $\log C$ | 0.70 | 0.60 | 0.48 | 0.30 | 0.18 | 0.00 | -0.10 | -0.22 | -0.30 | -0.40 | -0.52 |
| $\log N(>C)$ | 1.48 | 1.60 | 1.70 | 1.88 | 1.96 | 2.13 | 2.20 | 2.30 | 2.36 | 2.44 | 2.51 |
| $N(>C)$ | 30 | 40 | 50 | 77 | 92 | 134 | 158 | 200 | 230 | 276 | 327 |
| $\log N_E(>C)$ | 1.62 | 1.79 | 1.98 | 2.27 | 2.43 | 2.69 | 2.82 | 3.00 | 3.14 | 3.27 | 3.49 |
| $N_E(>C)$ | 41 | 62 | 96 | 185 | 270 | 493 | 665 | 992 | 1366 | 1845 | 3103 |
| $q(z)$ | 0.73 | 0.65 | 0.52 | 0.42 | 0.34 | 0.27 | 0.24 | 0.20 | 0.17 | 0.15 | 0.11 |
| $D=4$ | | | | | | | | | | | |
| $z$ | 0.23 | 0.33 | 0.55 | 0.78 | 1.05 | 1.39 | 1.59 | 1.92 | 2.26 | 2.54 | 3.36 |
| $\mathcal{L}_o h^2/10^{56}$ | 2.4 | 3.8 | 7.2 | 8.9 | 11.9 | 11.2 | 11.9 | 12.0 | 13.0 | 12.4 | 14.2 |
| $D=3$ | | | | | | | | | | | |
| $z$ | 0.17 | 0.22 | 0.35 | 0.48 | 0.63 | 0.80 | 0.90 | 1.06 | 1.24 | 1.36 | 1.67 |
| $\mathcal{L}_o h^2/10^{56}$ | 1.3 | 1.7 | 3.1 | 3.8 | 3.6 | 4.7 | 4.6 | 4.5 | 4.9 | 4.6 | 4.8 |
| $D=2$ | | | | | | | | | | | |
| $z$ | 0.12 | 0.16 | 0.26 | 0.35 | 0.44 | 0.56 | 0.63 | 0.74 | 0.83 | 0.91 | 1.12 |
| $\mathcal{L}_o h^2/10^{56}$ | 0.7 | 0.9 | 1.8 | 2.1 | 2.4 | 2.5 | 2.5 | 2.4 | 2.5 | 2.4 | 2.5 |
| $D=1$ | | | | | | | | | | | |
| $z$ | 0.10 | 0.12 | 0.20 | 0.28 | 0.34 | 0.42 | 0.48 | 0.56 | 0.62 | 0.68 | 0.83 |
| $\mathcal{L}_o h^2/10^{56}$ | 0.4 | 0.5 | 1.1 | 1.4 | 1.5 | 1.5 | 1.5 | 1.5 | 1.5 | 1.4 | 1.5 |
| $D=0$ | | | | | | | | | | | |
| $z$ | 0.08 | 0.10 | 0.17 | 0.23 | 0.27 | 0.34 | 0.39 | 0.45 | 0.50 | 0.55 | 0.66 |
| $\mathcal{L}_o h^2/10^{56}$ | 0.3 | 0.4 | 0.8 | 1.0 | 1.0 | 1.0 | 1.0 | 1.0 | 1.0 | 1.0 | 0.9 |

## Table Caption

**Table.1** Comparison of the cosmological unbounded standard candle assumption with comoving densities depending as $(1+z)^D$ for selected values of the count rate $C$. The case $D = 3$ describes the constant comoving density; $D = 4$ gives the case when the comoving density is increasing with $z$; $D < 3$ gives the cases when the comoving density is decreasing with $z$. If correct, one would obtain, for a given $D$, the same $\mathcal{L}_o$ at all $z$ or $C$. Here $C$ is in units of photons/(cm$^2$s), $\mathcal{L}_o$ is in units of photons/s.

## Figure Caption

**Fig.1.** Graph of $\log N(>C)$ versus $C$ as obtained from the publicly available BATSE 2B catalogue. Two straight lines with slopes -3/2 and -0.88 are also shown superimposed, representing the Euclidean limit and a particular power-law luminosity function distribution (see text).